\documentclass[12pt]{article}

\newcommand{\be}{\begin{equation}}
\newcommand{\ee}{\end{equation}}
\begin{document}
\title{ON THE ACTIVE GRAVITATIONAL MASS OF A NON--SPHERICAL SOURCE LEAVING  HYDROSTATIC EQUILIBRIUM}
\author{L. Herrera\thanks{Postal
address: Apartado 80793, Caracas 1080A, Venezuela.} \thanks{e-mail:
laherrera@telcel.net.ve}
, A. Di Prisco,
E. Fuenmayor\thanks{e-mail:efuenma@fisica.ciens.ucv.ve}
 \\
{\small Escuela de F\'{\i}sica, Facultad de Ciencias,}\\
{\small Universidad Central de Venezuela, Caracas, Venezuela.} \\
\\
}
\maketitle
\begin{abstract}
We obtain an expression for the active gravitational mass (Tolman) of  a source of the $\gamma$ metric, just after its departure from hydrostatic equilibrium, on a time scale of the
order of (or smaller than) the hydrostatic time scale. It is shown that for very compact sources, even  arbitrarily small departures from sphericity, produce significant decreasing
(increasing) in the values of active gravitational mass of collapsing (expanding) spheres, with respect to its value in  equilibrium, enhancing thereby the stability of the system.
\end{abstract}
\pagebreak
\section{Introduction}
One of the most remarkable aspect of general relativity, is the very special status that this theory confers to  the spherical symmetry. This situation is particularly well 
illustrated by the  Israel theorem.

Indeed, as is well known \cite{1},
the
only static and asymptotically-flat vacuum space-time possessing a
regular
horizon is the Schwarzschild solution. For all the others Weyl exterior
solutions \cite{2}, the physical components of the Riemann tensor
exhibit
singularities at $r=2m$.
 Therefore, it is intuitively
clear that for high gravitational fields, the evolution of sources of
Weyl
space-time should drastically differ from the evolution of spherical
sources \cite{Bel}. It is important to keep in mind that the sharp difference in
the
behaviour of both types of sources (for very high gravitational fields)
will
exist independently on the magnitude of multipole moments (higher than
monopole)
of the Weyl source. This is so because, as the source approaches the
horizon, any finite perturbation of the Schwarzschild space-time becomes
fundamentally different from any Weyl solution, even when the latter is
characterized by parameters whose values are arbitrarily close to those
corresponding to the spherical symmetry. This point has been stressed
before \cite{4}, but usually it has been overlooked.

Notwithstanding, spherical symmetry is a common assumption in the study of compact self-gravitating objects (white dwarfs, neutron stars, black holes
) , furthermore in the specific case of non-rotating black holes, spherical symmetry should be``absolute'', according to Israel 
theorem. Therefore it is pertinent to ask, how do small deviations from this assumption, related to any kind of perturbation 
(e.g. fluctuations of the stellar matter, external perturbations, etc), affect the dynamics of the system?.

In a recent work \cite{Herrera} it was shown that for a non--spherical source (even in the case of slight deviations from spherical symmetry), the speed of entering the collapse regime
decreases substantially, as compared with the exactly spherically symmetric case. Also, the sensitivity of the trajectories of test particles in the $\gamma$ spacetime, to small
changes of $\gamma$, for orbits close to $2m$, has been brought out \cite{HS}.

It is the purpose of this work to study further the behaviour of axysymmetric sources for very high
gravitational fields. This will allow us to put in evidence the role played by the non sphericity (however small) of the source,
on the outcome of evolution. To do so we shall obtain an expression for the active gravitational mass of the source, immediately after its departure from equilibrium. Here
``immediately'' means on a time scale
of the order of (or smaller than) hydrostatic time scale -see section 4 for more details. 

As initial configurations, we shall consider two interior metrics. These
ones
were found some years ago by Stewart {\it et al.} \cite{7}, following a
prescription given by Hern\'{a}ndez \cite{8} allowing to
obtain sources of Weyl space time, from known spherically
symmetric interior solutions.

The configurations to be considered are sources of the so-called
gamma
metric ($\gamma $-metric) \cite{9}, \cite{10}. This metric, which is
also
known as Zipoy-Vorhees metric \cite{11}, belongs to the family of Weyl's
solutions, and is continuously linked to the Schwarzschild space-time
through one of its parameters. The motivation for this choice is
twofold. On
one hand the exterior $\gamma $-metric corresponds to a solution of the
Laplace equation (in cylindrical coordinates) with the same singularity
structure as the Schwarzschild solution (a line segment \cite{9}). In
this
sense the $\gamma $-metric appears as the ``natural'' generalization of
Schwarzschild space-time to the axisymmetric case. On the other hand,
the
two interiors considered have reasonable physical properties
and
generalize important and useful sources of the Schwarzschild space-time,
namely the interior Schwarzschild solution (homogeneous density) and the
Adler solution \cite{12}. All this having been said, we would like to emphasize that our main goal here is not to model the behaviour of a specific type of compact object, but to
illustrate the effects of slight deviations from spherical symmetry, on the source.

As we shall see, the value of active gravitational mass of the source evaluated just after the departure from equilibrium, will be seriously affected by slight deviations from
spherical symmetry.

 On the other hand, we are well aware of the fact that the $\gamma $ metric is not the only
possible description for the exterior of a compact objetc and, of course, the two equations of state considered here do not exhaust 
the list of possible candidates for the equation of state of the stellar matter. However, in view of the properties of the $\gamma $ 
metric and the two equations of state considered here, mentioned above, it is fair to say that the case for the relevance of small 
deviations from spherical symmetry in the dynamics of compact objects has been further strengthened. This suggests that any conclusion on the 
structure and evolution of a compact object, derived on the assumption of spherical symmetry should be carefully   checked against
small deviations from that assumption, whenever the boundary of the source is close to the horizon.

\section{The Space-time}

\subsection{The exterior space-time}

As it has been mentioned above, our initial matter configuration is the
source
of an axially symmetric and static space-time ($\gamma $-metric).
In cylindrical
coordinates,
static axisymmetric solutions to Einstein equations are given by the
Weyl
metric \cite{2}
\begin{equation}
ds^2=e^{2\lambda }dt^2-e^{-2\lambda }\left[ e^{2\mu }\left( d\rho
^2+dz^2\right) +\rho ^2d\varphi ^2\right] ,  \label{1}
\end{equation}
with
\begin{equation}
\lambda _{,\rho \rho }+\rho ^{-1}\lambda _{,\rho }+\lambda _{,zz}=0
\label{2}
\end{equation}
and
\begin{equation}
\mu _{,\rho }=\rho \left(\lambda _{,\rho }^2-\lambda _{,z}^2\right)\qquad
\mu _{,z}=2\rho
\lambda _{,\rho }\lambda _{,z}.  \label{3}
\end{equation}
Observe that (\ref{2}) is just the Laplace
equation for $\lambda $ (in the Euclidean space).

The $\gamma $-metric is defined by \cite{9}
\begin{equation}
\lambda =\frac \gamma 2\ln \left[ \frac{R_1+R_2-2m}{R_1+R_2+2m}\right] ,
\label{4}
\end{equation}
\begin{equation}
e^{2\mu} =\left[ \frac{\left( R_1+R_2+2m\right) \left( R_1+R_2-2m\right)
}{%
4R_1R_2}\right] ^{\gamma ^2},  \label{5}
\end{equation}
where
\begin{equation}
R_1^2=\rho ^2+(z-m)^2\qquad R_2^2=\rho ^2+(z+m)^2.  \label{6}
\end{equation}
It is worth noticing that $\lambda ,$ as given by (\ref{4}), corresponds
to
the Newtonian potential of a line segment of mass density $\gamma/2 $ and
length $2m,$ symmetrically distributed along the $z$ axis. The
particular
case $\gamma =1,$ corresponds to the Schwarzschild metric.

It will be useful to work in Erez-Rosen coordinates \cite{11}, given by
\begin{equation}
\rho ^2=(r^2-2mr)\sin ^2\theta \qquad z=(r-m)\cos \theta ,  \label{7}
\end{equation}
which yields the line element as \cite{9}
\begin{equation}
ds^2=Fdt^2-F^{-1}\left\{ Gdr^2+Hd\theta ^2+\left( r^2-2mr\right) \sin
^2\theta d\varphi ^2\right\} ,  \label{8}
\end{equation}
where
\begin{eqnarray}
F &=&\left( 1-\frac{2m}r\right) ^\gamma ,  \label{8a} \\
&&  \nonumber \\
G &=&\left( \frac{r^2-2mr}{r^2-2mr+m^2\sin ^2\theta }\right) ^{\gamma
^2-1},
\label{8b}
\end{eqnarray}
and
\begin{equation}
H=\frac{\left( r^2-2mr\right) ^{\gamma ^2}}{\left( r^2-2mr+m^2\sin
^2\theta
\right) ^{\gamma ^2-1}}  \label{8c}
\end{equation}
Now, it is easy to check that $\gamma =1$ corresponds to the
Schwarzschild
metric.

The total mass of the source is \cite{9,10} $M=\gamma m,$ and the
quadrupole
moment is given by
\begin{equation}
Q=\frac \gamma 3m^3\left( 1-\gamma ^2\right) .  \label{9}
\end{equation}
So that $\gamma >1$ ($\gamma <1$) corresponds to an oblate (prolate)
spheroid.

\subsection{The interior space-time}

The metric within the matter distribution bounded by the surface
\begin{equation}
r=r_{\Sigma}  \label{10}
\end{equation}
is given by
\begin{eqnarray}
g_{tt} &=&f^{2\gamma }  \nonumber \\
g_{rr} &=&-f^{2(1-\gamma )}\Delta ^{\gamma ^2-2}\Sigma ^{1-\gamma ^2}
\nonumber \\
g_{\theta \theta } &=&-r^2f^{2\gamma (\gamma -1)}\Phi ^{1-\gamma^2}
\nonumber \\
g_{\varphi \varphi } &=&-r^2f^{2(1-\gamma )}\sin ^2\theta
\label{11}
\end{eqnarray}
where $f,$ $\Delta ,$ $\Sigma $ and $\Phi $ are functions whose specific
form depends on the model under consideration.

The two cases to be considered here are  reported in
\cite{7},
namely

\begin{enumerate}
\item  The modified constant density Schwarzschild solution
\begin{eqnarray}
f(r)&=&\frac 32\sqrt{1-\frac{r_{\Sigma}^2}{B^2}}-\frac 12\sqrt{1-\frac{r^2}{B^2}}
\label{12} \\
\nonumber \\
\Delta (r)&=&1-\frac{r^2}{B^2}  \label{13} \\
\nonumber \\
\Sigma (r,\theta )&=&1-\frac{r^2}{B^2}+\frac{r^4}{4B^4}\sin ^2\theta
\label{14} \\
\nonumber \\
\Phi (r,\theta )&=&f^2+\frac{r^4}{4B^4}V(r)\sin ^2\theta   \label{15}
\end{eqnarray}
with
\begin{equation}
V(r)=1+\frac 6r_{\Sigma}\left( r_{\Sigma}-r\right) ,  \label{16}
\end{equation}
and
\begin{equation}
B^2=\frac 3{8\pi \rho _{ss}},  \label{17}
\end{equation}
where $\rho _{ss}$ denotes the energy density in the spherically
symmetric
limit ($\gamma =1$)

\item  The modified Adler solution
\begin{eqnarray}
f(r)&=&A+Br^2  \label{18} \\
\nonumber \\
\Delta (r)&=&1+\frac{Cr^2}{\left( A+3Br^2\right) ^{2/3}}  \label{18a} \\
\nonumber \\
\Sigma (r,\theta )&=&1+\frac{Cr^2}{\left( A+3Br^2\right)
^{2/3}}+\frac{C^2r^4}{%
4\left( A+3Br^2\right) ^{4/3}}\sin ^2\theta   \label{19} \\
\nonumber \\
\Phi (r,\theta )&=&\left( A+Br^2\right) ^2+\frac{C^2r^4V(r)}{4\left(
A+3Br^2\right) ^{4/3}}\sin ^2\theta   \label{20} \\
\end{eqnarray}
with
\begin{equation}
V(r)=1+\frac 6r_{\Sigma}\left( 1-\frac{5m}{3r_{\Sigma}}\right) \left( 1-\frac mr_{\Sigma}\right)
^{-1}\left( r_{\Sigma}-r\right)   \label{21}
\end{equation}
and
\begin{eqnarray}
A&=&\frac{1-\frac{5m}{2r_{\Sigma}}}{\left(1-\frac{2m}{r_{\Sigma}}\right)^{1/2}}
\nonumber \\
B&=&\frac m{2r_{\Sigma}^3\left( 1-\frac{2m}{r_{\Sigma}}\right) ^{1/2}}  \nonumber \\
\nonumber \\
C&=&-\frac{2m\left( 1-\frac {m}{r_{\Sigma}}\right) ^{2/3}}{r_{\Sigma}^3\left( 1-\frac{2m}{r_{\Sigma}}\right)
^{1/3}}  \label{22}
\end{eqnarray}
\end{enumerate}

Before closing this section, two remarks are in order:

\begin{enumerate}
\item  Since we are considering the source described in (\ref{11}) as an
initial state, the time derivatives of functions $f,$ $\Delta ,$ $\Sigma$
and $\Phi $ will be in principle different from zero.

\item  Junction (Darmois) conditions are satisfied at the boundary $r=r_{\Sigma}$
-see \cite{7} for details.
\end{enumerate}

\section{The energy momentum tensor}

In order to give physical meaning to the components of the energy
momentum
tensor in coordinates ($t,$ $r,$ $\theta ,$ $\varphi $), we shall
develop a
procedure similar to that used by Bondi \cite{Bo} in his study of non
static
spherically symmetric sources. Thus, we introduce purely local Minkowski
coordinates ($\tau ,$ $x,$ $y,$ $z$) (alternatively one may introduce  a tetrad field associated with locally Minkowskian observers) defined by
\begin{eqnarray}
d\tau &=&f^\gamma dt  \label{31} \\
&&  \nonumber \\
dx &=&f^{1-\gamma }\Delta ^{-1+\gamma ^2/2}\Sigma ^{(1-\gamma ^2)/2}dr
\label{32} \\
&&  \nonumber \\
dy &=&rf^{\gamma (\gamma -1)}\Phi ^{(1-\gamma ^2)/2}d\theta  \label{33}
\\
&&  \nonumber \\
dz &=&r\sin (\theta )f^{1-\gamma }d\varphi .  \label{34}
\end{eqnarray}

Next, since we are assuming that our source does not dissipate energy,
then
the covariant components of the energy momentum tensor, as measured by a
local Minkowskian and comoving with the fluid observer, will be
\begin{equation}
\widehat{T}_{\mu \nu }=\left(
\begin{array}{cccc}
\rho & 0 & 0 & 0 \\
0 & p_{xx} & p_{xy} & 0 \\
0 & p_{yx} & p_{yy} & 0 \\
0 & 0 & 0 & p_{zz}
\end{array}
\right) ,  \label{35}
\end{equation}
where $\rho $ is the energy density and in general $p_{xx}\neq
p_{yy}\neq
p_{zz}$ and $p_{xy}=p_{yx}.$ We may write (\ref{35}) in the form
\begin{equation}
\widehat{T}_{\mu \nu }=\left( \rho +p_{zz}\right) \widehat{U}_\mu
\widehat{U}%
_\nu -p_{zz}\eta _{\mu \nu }+\left( p_{xx}-p_{zz}\right) \widehat{k}_\mu
\widehat{k}_\nu +\left( p_{yy}-p_{zz}\right) \widehat{l}_\mu
\widehat{l}_\nu
+2p_{xy}\widehat{k}_{(\mu }\widehat{l}_{\nu )},  \label{36}
\end{equation}
where $\eta _{\mu \nu }$ denotes the flat space-time metric and
\begin{eqnarray}
\widehat{U}_\mu &=&\left(
\begin{array}{cccc}
1, & 0, & 0, & 0
\end{array}
\right)  \label{37} \\
&&  \nonumber \\
\widehat{k}_\mu &=&\left(
\begin{array}{cccc}
0, & 1, & 0, & 0
\end{array}
\right)  \label{38} \\
&&  \nonumber \\
\widehat{l}_\mu &=&\left(
\begin{array}{cccc}
0, & 0, & 1, & 0
\end{array}
\right)  \label{39b}
\end{eqnarray}

The components of the energy-momentum tensor $T_{\mu \nu }$ in ($t,$
$r,$ $%
\theta ,$ $\varphi $) coordinates are linked to (\ref{36}) by
\begin{equation}
T_{\mu \nu }=\Lambda _\mu ^\alpha \Lambda _\nu ^\beta L_\alpha ^\gamma
L_\beta ^\delta \widehat{T}_{\gamma \delta },  \label{40}
\end{equation}
where the non vanishing components of $\Lambda _\mu ^\nu $ are
\begin{eqnarray}
\Lambda _t^\tau &=&f^\gamma  \label{43} \\
&&  \nonumber \\
\Lambda _r^x &=&f^{1-\gamma }\Delta ^{-1+\gamma ^2/2}\Sigma ^{(1-\gamma
^2)/2}  \label{44} \\
&&  \nonumber \\
\Lambda _\theta ^y &=&rf^{\gamma (\gamma -1)}\Phi ^{(1-\gamma ^2)/2}
\label{45} \\
&&  \nonumber \\
\Lambda _\varphi ^z &=&r\sin (\theta )f^{1-\gamma }.  \label{46}
\end{eqnarray}
 and the Lorentz matrices $L_\mu ^\nu
$
are given by
\begin{equation}
L_t^t=\Gamma \qquad L_i^t=L_t^i=-\Gamma w_i\qquad L_j^i=L_i^j=\delta
_j^i+%
\frac{(\Gamma -1)w_iw_j}{w^2},  \label{41}
\end{equation}
where
\begin{equation}
w^2=w_x^2+w_y^2\qquad \Gamma =\frac 1{\sqrt{1-w^2}},  \label{42}
\end{equation}
and $w_x$ and $w_y$ denote, respectively, the velocity of a fluid
element
along the $x$  and $y$ ($r$ and $\theta $) directions, as measured by our
local
Minkowskian observer as defined by (\ref{31})--(\ref{34}). Observe that we
are
considering the case $w_z=0,$ which means that the system preserves the
reflection symmetry (besides the axial symmetry).

Then, (\ref{40}) readily gives
\begin{eqnarray}
T_{tt} &=&f^{2\gamma }\Gamma ^2\left( \rho
+p_{xx}w_x^2+p_{yy}w_y^2+2p_{xy}w_xw_y\right)   \label{47} \\
&&  \nonumber \\
T_{tr} &=&-f\Delta ^{-1+\gamma ^2/2}\Sigma ^{(1-\gamma ^2)/2}\Gamma
\times
\nonumber \\
&&\left( \Gamma w_x\rho +p_{xx}w_x\Lambda _x+p_{yy}w_y\Lambda
+p_{xy}\left[
w_x\Lambda +w_y\Lambda _x\right] \right)   \label{48} \\
&&  \nonumber \\
T_{t\theta } &=&-rf^{\gamma ^2}\Phi ^{(1-\gamma ^2)/2}\Gamma \times
\nonumber \\
&&\left( \Gamma w_y\rho +p_{xx}w_x\Lambda +p_{yy}w_y\Lambda
_y+p_{xy}\left[
w_y\Lambda +w_x\Lambda _y\right] \right)   \label{49} \\
&&  \nonumber \\
T_{rr} &=&f^{2-2\gamma }\Delta ^{\gamma ^2-2}\Sigma ^{1-\gamma ^2}\times
\nonumber \\
&&\left( \Gamma ^2w_x^2\rho +p_{xx}\Lambda _x^2+p_{yy}\Lambda
^2+2p_{xy}\Lambda \Lambda _x\right)   \label{50} \\
&&  \nonumber \\
T_{r\theta } &=&rf^{(\gamma -1)^2}\Delta ^{-1+\gamma ^2/2}\Phi
^{(1-\gamma
^2)/2}\Sigma ^{(1-\gamma ^2)/2}\times  \nonumber \\
&&\left( \Gamma ^2w_xw_y\rho +\Lambda \left[ p_{xx}\Lambda
_x+p_{yy}\Lambda
_y\right] +p_{xy}\left[ \Lambda ^2+\Lambda _x\Lambda _y\right] \right)
\label{51} \\
&&  \nonumber \\
T_{\theta \theta } &=&r^2f^{2\gamma (\gamma -1)}\Phi ^{1-\gamma
^2}\times
\nonumber \\
&&\left( \Gamma ^2w_y^2\rho +p_{xx}\Lambda ^2+p_{yy}\Lambda
_y^2+2p_{xy}\Lambda \Lambda _y\right)   \label{52} \\
&&  \nonumber \\
T_{\varphi \varphi } &=&r^2f^{2(1-\gamma )}\sin ^2(\theta )p_{zz},
\label{53}
\end{eqnarray}
with
\begin{eqnarray}
\Lambda  &\equiv &\frac{\left( \Gamma -1\right) w_xw_y}{w^2},
\label{53a} \\
&&  \nonumber \\
\Lambda _x &\equiv &1+\frac{\left( \Gamma -1\right) w_x^2}{w^2},
\label{53b}
\\
&&  \nonumber \\
\Lambda _y &\equiv &1+\frac{\left( \Gamma -1\right) w_y^2}{w^2},
\label{53c}
\end{eqnarray}

So that
\begin{equation}
{T}_{\mu \nu }=\left( \rho +p_{zz}\right) {U}_\mu
{U}%
_\nu -p_{zz} g_{\mu \nu }+\left( p_{xx}-p_{zz}\right) {k}_\mu
{k}_\nu +\left( p_{yy}-p_{zz}\right) {l}_\mu
{l}_\nu
+2p_{xy}{k}_{(\mu }{l}_{\nu )},  \label{Tmn}
\end{equation}
where $U_\mu$, $k_\mu$ and $l_\mu$ are obtained
after applying the boost velocity (\ref{41}) and  (\ref{43})--(\ref{46}) to (\ref{37})--(\ref{39b}),
\begin{equation}
U_\mu =\Gamma \left(
\begin{array}{cccc}
f^\gamma , & -w_xf^{1-\gamma }\Delta ^{-1+\gamma ^2/2}\Sigma ^{(1-\gamma
^2)/2}, & -w_yrf^{\gamma (\gamma -1)}\Phi ^{(1-\gamma ^2)/2}, & 0
\end{array}
\right) ,  \label{53e}
\end{equation}
\begin{equation}
k_\mu =\left(
\begin{array}{cccc}
-\Gamma w_xf^\gamma , & f^{1-\gamma }\Delta ^{-1+\gamma ^2/2}\Sigma
^{(1-\gamma ^2)/2}\Lambda _x, & rf^{\gamma (\gamma -1)}\Phi ^{(1-\gamma
^2)/2}\Lambda , & 0
\end{array}
\right) ,  \label{53f}
\end{equation}
\begin{equation}
l_\mu =\left(
\begin{array}{cccc}
-\Gamma w_yf^\gamma , & f^{1-\gamma }\Delta ^{-1+\gamma ^2/2}\Sigma
^{(1-\gamma ^2)/2}\Lambda , & rf^{\gamma (\gamma -1)}\Phi ^{(1-\gamma
^2)/2}\Lambda _y, & 0
\end{array}
\right) .  \label{53g}
\end{equation}

In the static case $w_x =w_y=\Lambda=0$ and $\Gamma=\Lambda_x=\Lambda_y=1$. The same holds after the system departs from equilibrium, on a time scale of the order of (or smaller than)
the hidrostatic time scale (see Section 2.4).
\section{The Tolman mass and departure from equilibrium}
We shall now find an expression for the Tolman mass of the source described in the previous section, an then we shall evaluate it just after its departure from equilibrium, on a
time scale of the order of (or smaller than) the hydrostatic time scale. Also, since we are interested in the effects produced by  small deviations from spherical symmetry, it  will be
assumed further that the source is quasi--spherical, and therefore we shall put
$\gamma=1+\epsilon$ and will neglect terms of order $O(\epsilon^2)$ or smaller.
\subsection{The Tolman mass}

The Tolman mass is given by \cite{To}

\begin{eqnarray}
m_T &=& \int^{r_{\Sigma}}_{0}\int^{\pi}_{0}\int^{2\pi}_{0}{\sqrt{-g}(T^0_0 - T^1_1 -T^2_2 - T^3_3) dr d\theta
d\phi} \nonumber \\
&+&
\frac{1}{8\pi}\int\int\int{\sqrt{-g} g^{\alpha\beta}
\frac{\partial}{\partial t}\left[\frac{\partial{\cal
L}}{\partial(\partial(\sqrt{-g}g^{\alpha\beta})/\partial t)}\right] dr
d\theta d\phi}
\label{def}
\end{eqnarray}

\noindent
where $\cal L$ denotes the usual gravitational lagrangian density
(eq.(10) in \cite{To}). Although Tolman's formula was introduced
as a measure of the total energy of the system, with no commitment
to its localization, we shall define the mass for any value of $r$, smaller or equal to $r_{\Sigma}$, as

\begin{eqnarray}
m_T &=& \int^{r}_{0}\int^{\pi}_{0}\int^{2\pi}_{0}{\sqrt{-g}(T^0_0 - T^1_1 -T^2_2 - T^3_3) dr d\theta
d\phi} \nonumber \\
&+&
\frac{1}{8\pi}\int\int\int{\sqrt{-g} g^{\alpha\beta}
\frac{\partial}{\partial t}\left[\frac{\partial{\cal
L}}{\partial(\partial(\sqrt{-g}g^{\alpha\beta})/\partial t)}\right] dr
d\theta d\phi}.
\label{Tolin}
\end{eqnarray}

\noindent
This extension of the global concept of energy to a local level
\cite{Coo} is suggested by the conspicuous role played by
$m_T$ as the ``active gravitational mass''.

Indeed, it can be easily shown \cite{Gro}
that the gravitational acceleration $a$ of a test particle,
instantaneously at rest in a static gravitational field, as measured
with standard rods and coordinate clock is given by

\begin{equation}
a  = - \frac{m_T}{r^2}.
\label{a}
\end{equation}

\noindent
A similar conclusion can be obtained by inspection of the equation of hydrostatic equlibrium (TOV)
(in the static or quasi-static, spherically symmetric case case) \cite{Lig}.

 Even though these properties of Tolman's definition of mass are only valid in the spherically symmetric and static case, it is reasonable to assume that for small deviations from these
conditions, the same role of active gravitational mass may be assigned to our expression.

\noindent
Let us now evaluate expression (\ref{Tolin}). After some lengthy calculations one finds:
\begin{eqnarray}
m_T &=& 2\pi \int^{r}_{0}\int^{\pi}_{0}{r^2 \sin{\theta} f^{\gamma^2 -2\gamma +2}
\Delta^{\gamma^2/2 - 1}
\Sigma^{(1-\gamma^2)/2}
\Phi^{(1-\gamma^2)/2}}\nonumber \\
&\times& \left[\Gamma^2 \rho(1 + \omega_{x}^2 + \omega_{y}^2) 
+ p_{xx} (\Gamma^2 \omega_{x}^2 + \Lambda_x^2 + \Lambda^2) 
+ p_{yy} (\Gamma^2 \omega_{y}^2 + \Lambda_y^2 +
\Lambda^2)\right.\nonumber \\
&& \left.  + p_{zz} 
+ 2 p_{xy} (\Gamma^2 \omega_{x} \omega_{y} + \Lambda(\Lambda_x +
\Lambda_y))\right] dr d\theta
\nonumber \\ 
&+&
\frac{1}{4} \int^{r}_{0}\int^{\pi}_{0}{r^2 \sin{\theta} f^{\gamma^2 -4\gamma +2}
\Delta^{\gamma^2/2 - 1}
\Sigma^{(1-\gamma^2)/2}
\Phi^{(1-\gamma^2)/2}} \nonumber \\
&\times&\left[
2 (2 - 3\gamma + \gamma^2) \frac{\ddot f}{f}
+ (\gamma^2 - 2) \frac{\ddot \Delta}{\Delta}
+ (1 - \gamma^2) \frac{\ddot \Phi}{\Phi}
+ (1 - \gamma^2) \frac{\ddot \Sigma}{\Sigma}\right.\nonumber \\
&& + 2\gamma (1 - \gamma^2) (-\gamma^2 +2\gamma -3)
\left(\frac{\dot f}{f}\right)^2
+ \frac{1}{2} (\gamma^2-2) (\gamma^2-4)
\left(\frac{\dot \Delta}{\Delta}\right)^2 \nonumber\\
&&- \frac{1}{2} (1-\gamma^2) (\gamma^2+1)
\left(\left(\frac{\dot \Phi}{\Phi}\right)^2 + \left(\frac{\dot
\Sigma}{\Sigma}\right)^2\right)
+ (\gamma^2 -2) (2-3\gamma) \frac{\dot f}{f} \frac{\dot \Delta}{\Delta} 
\nonumber \\
&& + (1-\gamma^2) (2-3\gamma) \frac{\dot f}{f} \frac{\dot
\Sigma}{\Sigma} + \gamma (1-\gamma^2) (2\gamma-3) \frac{\dot f}{f}
\frac{\dot
\Phi}{\Phi}\nonumber\\
&& \left. + (1-\gamma^2) (\gamma^2-2) \frac{\dot \Delta}{\Delta}
\frac{\dot
\Sigma}{\Sigma}
\right] dr d\theta
\label{mta}
\end{eqnarray}

We shall now evaluate this last expression at the very moment the system leaves the equilibrium, and for small non--sphericity.

\subsection{Departure from equilibrium}

Let us now consider that our  source,
 once submitted to perturbations, departs from equilibrium without
dissipation. We shall then evaluate the system after such departure, on a
time
scale such that $w_x$ and $w_y$ remain vanishingly small, whereas their
time
derivatives though small, will be different from zero.

Thus, just after leaving the equilibrium, the following conditions hold
\begin{equation}
w_x=w_y=w_{x,i}=w_{y,i}\simeq 0,\qquad (i=r,\theta ,\varphi )
\label{58}
\end{equation}
\begin{equation}
\dot w_{x},\ \dot w_{y}\neq 0\qquad (\mbox{small})  \label{59}
\end{equation}
where dots denote derivative with respect to $t$.

From now on, unless otherwise stated, all equations are evaluated
at the moment the system starts to deviate from equilibrium.

Then from (\ref{47})--(\ref{53}), we obtain using (\ref{58})

\begin{equation}
T_{t \theta} = T_{t r} = 0
\label{T0}
\end{equation}
which implies, because of field equations (see \cite{Herrera} for details)
\begin{equation}
\dot \Delta = \dot f = \dot \Sigma = \dot \Phi = 0
\label{der0}
\end{equation}
where for simplicity we write $0$ for ${\cal O}(\omega)$ (as we shall do
hereafter).

Obviously, spatial derivatives of the above quantities will be also
vanishingly small on the time scale under consideration.

Also, from evaluation of the $t$-component of the conservation law $T_{\nu;\mu}^{\mu}=0$, we get (see \cite{Herrera}).
\begin{equation}
\dot \rho=0  \label{62}
\end{equation}

Then, in the approximation above, the expression for the Tolman mass becomes
\begin{eqnarray}
m_T &=& 2\pi \int^{r}_{0}\int^{\pi}_{0}{r^2 \sin{\theta} f^{(\gamma^2 -2\gamma + 2)}
\Delta^{(\gamma^2/2 - 1)}
\Sigma^{(1-\gamma^2)/2}
\Phi^{(1-\gamma^2)/2}}
\nonumber \\
&\times& {(T^0_0 - T^1_1 -T^2_2 - T^3_3) dr d\theta d\phi}
\nonumber \\ 
&+&
\frac{1}{4}\int^{r}_{0}\int^{\pi}_{0}{r^2 \sin{\theta} f^{(\gamma^2 -4\gamma + 2)}
\Delta^{(\gamma^2/2 - 1)}
\Sigma^{(1-\gamma^2)/2}
\Phi^{(1-\gamma^2)/2}}
\nonumber \\ 
&\times& {\left[2(\gamma^2 -3\gamma + 2)\frac{\ddot f}{f} 
+(\gamma^2 - 2) \frac{\ddot\Delta}{\Delta} + (1 -
\gamma^2)\left(\frac{\ddot\Sigma}{\Sigma} + \frac{\ddot
\Phi}{\Phi}\right)\right]}  
{drd\theta d\phi}\nonumber\\
\label{mt}
\end{eqnarray}

We shall now write $\gamma=1+\epsilon$, then assuming that non--sphericity is small (quasi--spherical approximation), the expression above reads:
\begin{eqnarray}
m_T &=& 2\pi \int^{r}_{0}\int^{\pi}_{0}{r^2 \sin{\theta} f
\Delta^{(\epsilon-1/2)}
\Sigma^{-\epsilon}
\Phi^{-\epsilon}
\times (T^0_0 - T^1_1 -T^2_2 - T^3_3) dr d\theta d\phi}
\nonumber \\ 
&+&
\frac{1}{4}\int^{r}_{0}\int^{\pi}_{0}{r^2 \sin{\theta} f^{-(1+2\epsilon)}
\Delta^{(\epsilon-1/2)}
\Sigma^{-\epsilon}
\Phi^{-\epsilon}}
\nonumber \\ 
&\times& \qquad {\left[- \frac{\ddot\Delta}{\Delta} + 2 \epsilon
\left(- \frac{\ddot f}{f} + \frac{\ddot\Delta}{\Delta} -
\frac{\ddot\Sigma}{\Sigma} - \frac{\ddot\Phi}{\Phi} \right)\right]
dr d\theta d\phi} 
\nonumber\\
\label{mtse}
\end{eqnarray}

Within our two approximations (quasi--spherical and (\ref{58})--(\ref{der0})) it follows from (\ref{47})--(\ref{53c}) that:
\be
T^0_0 - T^1_1 -T^2_2 - T^3_3 = \rho + p_{xx} + p_{yy} + p_{zz}
\label{TI}
\ee
Also within these approximations, the time derivative of the $tr$ component of the Einstein equations $G_{rt} = - 8 \pi T_{rt}$, yields
\begin{eqnarray}
&-& \frac{1}{r} \frac{\ddot\Delta}{\Delta} + 2 \epsilon \left[
\frac{\ddot\Phi'}{2\Phi} + \frac{\Phi'}{2\Phi}
\frac{\ddot\Delta}{2\Delta} 
+ \frac{1}{r} \left(\frac{\ddot\Delta}{\Delta} -
\frac{\ddot\Sigma}{\Sigma} - \frac{\ddot f}{f} + \frac{\ddot\Phi}{2\Phi} 
\right) - \frac{\ddot\Phi}{2\Phi} \left(\frac{f'}{f} +
\frac{\Phi'}{\Phi}  \right)
\right]
\nonumber \\
&=& 8\pi f \Delta^{(\epsilon - 1/2)}\Sigma^{-\epsilon} \left[
\dot\omega_x (\rho + p_{xx}) + \dot\omega_y p_{xy}\right].
\label{Ep}
\end{eqnarray}
Feeding back (\ref{TI}) and (\ref{Ep}) into (\ref{mtse}), one obtains after some rearrangements
\begin{eqnarray}
m_T &=& 2\pi \int^{r}_{0}\int^{\pi}_{0}{r^2 \sin{\theta} f
\Delta^{(\epsilon-1/2)}
\Sigma^{-\epsilon}
\Phi^{-\epsilon}
\times (\rho + p_{xx} + p_{yy} + p_{zz}) dr d\theta}
\nonumber \\ 
&+&
2\pi \int^{r}_{0}\int^{\pi}_{0}{r^3 \sin{\theta} f^{-2\epsilon}
\Delta^{(2\epsilon-1)}
\Sigma^{-2\epsilon}
\Phi^{-\epsilon}\left[
\dot\omega_x (\rho + p_{xx}) + \dot\omega_y p_{xy}\right] dr d\theta} 
\nonumber \\ 
&-& \frac{3}{4} \epsilon \int\int{r^2 \sin{\theta} f^{-(1+2\epsilon)}
\Delta^{(\epsilon-1/2)}
\Sigma^{-\epsilon}
\Phi^{-\epsilon} \frac{\ddot \Phi}{\Phi}dr d\theta}
\nonumber \\
&-& \frac{1}{4} \epsilon \int\int{r^3 \sin{\theta} f^{-(1+2\epsilon)}
\Delta^{(\epsilon-1/2)}
\Sigma^{-\epsilon}
\Phi^{-\epsilon}}\nonumber \\
&\times& 
{\left[
\frac{\ddot\Phi'}{2\Phi} + \frac{\Phi'}{2\Phi}
\frac{\ddot\Delta}{2\Delta} 
+ \frac{1}{r} \left(\frac{\ddot\Delta}{\Delta} -
\frac{\ddot\Sigma}{\Sigma} - \frac{\ddot f}{f} + \frac{\ddot\Phi}{2\Phi} 
\right) - \frac{\ddot\Phi}{2\Phi} \left(\frac{f'}{f} +
\frac{\Phi'}{\Phi}  \right)
\right] dr d\theta}\nonumber \\
\label{mtf}
\end{eqnarray}
\subsection{The spherically symmetric case}
Before entering into the discussion of (\ref{mtf}), it is quite instructive to analyze the spherically symmmetric situation. In this case one easily obtains from (\ref{mtf})
\begin{eqnarray}
m_T &=& 4\pi \int^{r}_{0}{r^2 f
\times (\rho + p_{xx} + p_{yy} + p_{zz}) dr}
\nonumber \\ 
&+&
4\pi \int^{r}_{0}{r^3 
\Delta^{-1}\left[
\dot\omega_x (\rho + p_{xx})\right] dr}
\label{mtfss}
\end{eqnarray}
with $p_{yy}=p_{zz}$.
Now, the first term in (\ref{mtfss}) correspond to the active gravitational mass of the core of radius $r$ interior to $r_{\Sigma}$ before the system leaves the equilibrium. Let us now
assume that the system leaves the equilibrium and starts to collapse (expand), then $\dot\omega_x <0$ ($\dot\omega_x >0$) decreasing (increasing) the value of the Tolman mass with
respect to its value at equilibrium. Thus the second term in (\ref{mtfss}) tends to stabilize the system, opposing any departure from equilibrium (contraction or expansion). This result
was already known \cite{inhomo}. 

Let us now get back to the non--spherical case to see how small non--sphericities change the whole picture.

\subsection{The non--spherical case}
Let us now turn back to (\ref{mtf}), to infer what happens
in the
non-spherical case (i.e. $\varepsilon \not= 0$), even though $\varepsilon<<1.$
Then neglecting higher order terms on $\varepsilon$, and keeping only the leading terms, we get
\begin{eqnarray}
m_T &=& 2\pi \int^{r}_{0}\int^{\pi}_{0}{r^2 \sin{\theta} f
\Delta^{(\epsilon-1/2)}
\Sigma^{-\epsilon}
\Phi^{-\epsilon}
\times (\rho + p_{xx} + p_{yy} + p_{zz}) dr d\theta}
\nonumber \\ 
&+&
2\pi \int^{r}_{0}\int^{\pi}_{0}{r^3 \sin{\theta} f^{-2\epsilon}
\Delta^{(2\epsilon-1)}
\Sigma^{-2\epsilon}
\Phi^{-\epsilon}\left[
\dot\omega_x (\rho + p_{xx}) \right] dr d\theta} 
\nonumber \\
\label{och}
\end{eqnarray}

In order to extract more information from (\ref{och}) it is necessary to
specify the source under consideration.
We shall use the two configurations mentioned in section 2.
In both cases, $f$ vanishes at the central region for $r_{\Sigma}>2m$. Thus, in the Schwarzschild-like
models we have $f(0)=0$, if $2m/r_{\Sigma} = 8/9$.
Since we know that any spherically symmetric static configuration with
constant $\rho_{ss}$ and locally
isotropic pressure should satisfy the constraint $n \equiv 2M/r_{\Sigma}<8/9$ we may
assume that the system leaves the equilibrium
for values of $n$ close to $8/9$.
Then, an inspection of (\ref{och}) shows that (if $\varepsilon>0$), for
values of $n$ approaching
$8/9$, the second term incresases due to the factor $f^{-2\epsilon}$.

We may now use Einstein equations to 
elucidate that for negative values of $\varepsilon$ the system may be 
unphysical. 
If $n$ takes values close to the limit allowed by the model
(8/9 for Schwarzschild-type model and 4/5 for Adler-type model),
the critical values ($f\rightarrow 0$)
appear close to $r=0$. Outside of this region the system is basically 
composed by an spherical incompressible fluid plus a perturbation in 
$\varepsilon$. 
Thus, the physical or unphysical character of the model is determined 
by its behaviour close to these two limits.  

The energy density, in the limit $r\rightarrow 0$,  for 
the Schwarzschild-type model and Adler-type model is given 
by expresions see (\cite{Herrera} for details)

\begin{equation}
\rho = \frac{m}{2\pi r_{\Sigma}^3}f^{2\varepsilon}\left[ \frac{3}{2}+ 
\varepsilon \left( 1+ \frac{1}{3\sqrt{1-\frac{2m}{r_{\Sigma}}}-1}\right)
\right]
\label{schdensid}
\end{equation}

and

\begin{equation}
\rho = \frac{3m}{4\pi r_{\Sigma}^3 \left( 1- 2m/r_{\Sigma}\right)^{1/3}} 
\left(\frac{1-5m/2r_{\Sigma}}{\sqrt{1-2m/r_{\Sigma}}}\right)^{2\varepsilon-2/3}
\left[ \left(1-m/r_{\Sigma}\right)^{2/3}+ 
\frac{\varepsilon}{\left(1-5m/2r_{\Sigma}\right)^{1/3}}\right]
\label{adldensid}
\end{equation}
 respectively.   
From these ones, it is easy to show that if $\varepsilon<0$, 
the energy density becomes negative 
as the system approaches to $n\rightarrow 8/9$ (Schwarzschild case)
or $n\rightarrow 4/5$ (Adler case) and positive for $\varepsilon \ge 0$.
Therefore, in both cases, positive energy conditions impose $\varepsilon \ge 0$,
and for very compact objects (close to the limit allowed by the model) 
the ``stabilizying''term in the expression for the active gravitational mass substantially increases in the central region, close to, but before, the maximum allowed value of $n$.

\section{Conclusions}
We have seen so far that (as expected), for high gravitational fields,
important differences appear in the way that  the spherical and  non-spherical
systems depart from equilibrium. This conclusion being true even for small non-sphericity.

In the two models considered above, a factor multiplying the stabilizying 
term in the expression for the active gravitational mass, brings out those differences. Although for the two examples
considered
here, we have considered $\varepsilon >0,$ it is obvious that
there
exist models with $\varepsilon <0,$ in which case the stabilizying term may become very small, leading to highly unstable
situations.
In fact, it is worth noticing that important differences between the two
cases ($\varepsilon >0$ and $\varepsilon <0$) appear also in the
behaviour of the exterior $\gamma $-metric (\cite{9}, \cite{10})

Thus, on the basis of presented results we may conclude that whatever
the
model and the sign of $\varepsilon $ would be, any source of Weyl metric
would evolve quite differently from the corresponding spherical source, as
critical values of $n$ are considered.

In particular, in the examples examined,  the departure from equilibrium appears to be affected by a sharp modification  in the
stabilyzying  term of the active gravitational mass,
 as $n$ approaches its maximum allowed value.
This
change makes the system more stable, hindering its departure from
equilibrium.

Before concluding, the following remarks are in order:
\begin{enumerate}
\item
Observe that all results above are valid only on a time scale of the order of (or smaller than) hydrostatic time scale. Within this time scale, the general form of the metric and
 the components of the energy--momentum tensor, is the same as in the static case (of course in the former case time dependence of variables has to be taken into account when taking
time derivatives of such  components)
\item
In [7] it is shown that the two interiors described in Section 2.2, match smoothly (on $r_{\Sigma}$) with the $\gamma$ metric, in the static case. That this is also the case after the
system departs from equilibrium (on a time scale of the order of, or smaller than, hydrostatic time scale) is apparent from (\ref{der0}) and comments after equation (57) and in the
point 1 above. Of course in the fully dynamic case (beyond the hydrostatic time scale), (\ref{der0}) does not longer hold, and matching conditions have to be established. But this is
out of the scope of this paper.
\end{enumerate}

\section*{acknowledgements}
 LH acknowledges financial assistance under grant
BFM2000-1322 (M.C.T. Spain).

\end{document}